\documentclass[12pt]{article}
\pdfoutput=1
\usepackage[margin=2cm]{geometry}
\usepackage{comment}
\usepackage{amsmath,amssymb,extarrows,mathtools,graphicx,subfigure,setspace}
\usepackage{cite}
\usepackage{slashed}
\usepackage{color}
\makeatother

\newcommand{\be}{\begin{equation}}
\newcommand{\bea}{\begin{eqnarray}}
\newcommand{\eea}{\end{eqnarray}}
\newcommand{\ba}{\begin{array}}
\newcommand{\ea}{\end{array}}
\newcommand{\ee}{\end{equation}}
\newcommand{\bes}{\begin{equation*}}
\newcommand{\beas}{\begin{eqnarray*}}
\newcommand{\eeas}{\end{eqnarray*}}
\newcommand{\bas}{\begin{array*}}
\newcommand{\eas}{\end{array*}}
\newcommand{\ees}{\end{equation*}}

\numberwithin{equation}{section}
\begin{document}
\onehalfspacing
\noindent
\begin{titlepage}
\hfill
\vbox{
    \halign{#\hfil         \cr
           IPM/P-2012/030 \cr
                      } 
      }  
\vspace*{20mm}
\begin{center}
{\Large {\bf Probing Fractionalized Charges}\\
}

\vspace*{15mm}
\vspace*{1mm}
{Mohsen Alishahiha$^a$  and Davood Allahbakhshi$^b$  }

 \vspace*{1cm}
{\it $^a$School of Physics , \\ $^b$ School of Particles and Accelerators,\\  
 Institute for Research in Fundamental Sciences (IPM)\\
P.O. Box 19395-5531, Tehran, Iran \\ }

\vspace*{.4cm}

{E-mails: {\tt allahbakhshi@ipm.ir, alishah@ipm.ir}}%

\vspace*{2cm}
\end{center}
\begin{abstract}
Inspired by the holographic entanglement entropy, for geometries with non-zero  abelian charges, we define a quantity  which is sensitive to the background  charges. 
One observes that there is a critical charge below that the system is  mainly described by the metric and the effects of the background charges are just via
metric's components. While for charges above the critical one the background gauge field plays an essential role. This, in turn,  might be used to  define an 
order parameter to probe phases of a system with fractionalized charges.

\end{abstract}
\end{titlepage}
\section{Introduction}

In application of AdS/CFT correspondence  \cite{M:1997} to condensed matter physics, one typically is
interested in a gravity dual which describes a system at finite temperature and density. Following  
\cite{Witten:1998zw} a natural guess for the dual gravity would be a  charged black hole. The existence of the charged horizon would result to a dual theory at finite temperature and finite density.

We, note, however that this is not the only way to construct a gravity model whose dual theory
is  a system at finite density. Indeed finite density holographic duals may be obtained 
by two, rather distinctive, ways. Actually  the asymptotic electric flux-to be identified with the
chemical potential at the boundary theory- may be supported by either non-zero charges from behind an event horizon, or charged matter in the bulk geometry. If we are interested in a phase
with  unbroken $U(1)$ global group, the matter filed in the bulk is charged fermions (see for
example \cite{Hartnoll:2011}).

Of course one can distinguish between these two cases due to the fact that in the first 
case (fractionalized phase)  the charge density is of oder of $N^2$ while in the second case
(mesonic phase) it is of order  ${\cal O}(N^0)$, where 
$N$ is the number of degrees of freedom (the number of color for $U(N)$ gauge theory).  
Alternatively, when the $U(1)$ is unbroken, the fractionalized phase may also be identified by
the violation of the Luttingger theorem \cite{{SSV:2002},{Sachdev:2010um},{Huijse:2011hp}}.

Since the  charge density of a system may be originated by both from behind an even horizon and a charged matter,  it could be in different phases depending  on the  
origin of the asymptotic flux. To classified  possible phases an order parameter has been introduced in \cite{Hartnoll:2012ux}. This order parameter at leading order is essentially the 
holographic entanglement entropy with taking into account the electric fluxes through the hypersurface of 
holographic  entanglement entropy.  In the present  paper we would like to introduce an order parameter which may probe a system with the fractionalized charges.  

To proceed, let us consider a $d+2$ dimensional Einstein-Dilaton-Maxwell theory whose action, 
in minimal form, may be written as follows
\be\label{action}
I=\frac{1}{16\pi G_{d+2}}\int d^{d+2}x\;\sqrt{-g}\left[{\bf R}-\frac{1}{2}(\partial\phi)^2+V(\phi)
-\frac{1}{4}\sum_{i=1}^ne^{\lambda_i\phi}F_i^2\right].
\ee
where $G_{d+2}$  is the $d+2$  dimensional Newton constant and $\lambda_i$'s are parameters
of model. This is, indeed, a typical action 
we get from compactification of low energy effective action of string theory. Of course this is the
case for particular values of the parameters $\lambda_i$ and a specific  form of the potential.
Nevertheless in what follows we will not restrict  ourselves to these particular values.

A generic solution of the equations of motion of the above action could be a charged black hole
(brane) with non-trivial dilaton profile.  We may assume that the background solution to be an
asymptotically locally $AdS_{d+2}$. Therefore the solution may provide a gravitational 
dual for a $d+1$ dimensional theory at finite charge and temperature with a  UV fixed point. 

The gravity description may be used to extract certain information about the dual field theoy.
In particular one may study certain non-local observables. 
%
Prototype 
examples include holographic entanglement entropy\cite{RT:2006PRL}  and Wilson loop\cite{{Rey:1998ik},{Maldacena:1998im}}. In both cases
the gravitational dual is found useful for extracting  the corresponding information. In both cases the 
problem  reduces to minimizing an area of a hypersurface in the bulk gravity.
Actually motivated by these quantities we would like to define a similar object which is also sensitive to the 
background gauge field. 

We note, however, that  since typically we are interested in backgrounds with electric field it is not appropriate to work with fixed time as one does for 
the holographic entanglement entropy. In other words it would be  more natural to consider the geometric entropy\cite{{Fujita:2008zv},{Bah:2008cj}} which is defined as follows.
To be specific consider a finite temperature four dimensional quantum field theory on $S^1\times S^3$. The
metric of $S^3$ sphere may be parametrized as follows  
\be
d\Omega_d^2=d\theta+\sin^2\theta (d\psi^2+\sin^2\psi\;d\phi^2),
\ee
with $0\leq \theta,\psi\leq \pi$ and $0\leq \phi\leq 2\pi$.

Let us change the periodicity of $\phi$ into $0\leq \phi\leq 2\pi k$ which results
to conical singularities at $\psi=0,\pi$ for $k\neq 1$ with the deficit angle
$2\pi(1-k)$. Let us denote by $Z[k]$ the partition function of the theory on this singular space.
Then one may define a density matrix as follows
\be\label{pf}
{\rm Tr}\rho^k=\frac{Z[k]}{(Z[1])^k},
\ee
where $Z[1]$ is the partition function of theory on $S^1\times S^3$. Using the  definition of 
von-Neumann entropy, the geometric entropy is defined by
\be
S_G=-{\rm Tr}(\rho\log \rho)=-\partial_k\log\left( \frac{Z[k]}{(Z[1])^k}\right)\bigg\vert_{k=1}.
\ee
Restricting to a subsystem one can also define a reduced density matrix. Of course  
it is clear that the corresponding entropy  is different from the entanglement entropy,
though it may be related to it by a double Wick rotation.

 From gravity point of view it  is essentially similar to the entanglement entropy 
where one should minimize a codimension two hypersurface in the bulk. Though in the 
present case  one  considers a hypersurface with a spatial direction fixed.  Indeed  
to compute  the geometric entropy one usually  utilizes a double Wick rotation to promote a spatial direction to time direction. Of course as far as the computations in the gravity side are
concerned it is not necessary to do that.

Now consider a codimension two hypersurface in the bulk\footnote{In what 
follows we use a notation in which the bulk coordinates are given by  $x^\mu=(t,r,x^i) $ for $i=1,\cdots, d$.} parametrized by
coodrinates $\xi_a$ for $a=1,\cdots, d$. Then one may define  two natural quantities:
the induced metric and  the pull back of the gauge field on the world volume of 
the hypersurface which are given by 
\be
{\tilde g}_{ab}=\frac{\partial x^\mu}{\partial \xi_a}\frac{\partial x^\nu}{\partial \xi_b}g_{\mu\nu},
\;\;\;\;\;\;\;\;\;\;\;\;\;\;\;\;F^i_{ab}=\frac{\partial x^\mu}{\partial \xi_a}\frac{\partial x^\nu}{\partial \xi_b}F^i_{\mu\nu},
\;\;\;\;\;\;\;{\rm for}\;x^d={\rm fixed}
\ee
The geometric entropy can be defined in terms of the induced metric as $S_G= \int d^d\xi \sqrt{\det(
\tilde{g})}$ when the area of the hypersurface is minimized. On the other hand  motivated by 
DBI action in the string theory it is natural to  define the following quantity\footnote{In general one could have put  free
parameters in front of each $F_{ab}^i$'s in the square root and therefore one has an $n$-parameter family object. We would like to thank D. Tong for suggesting such a possibility.}
\bea\label{area}
{\Gamma}=\frac{1}{G_{d+2}}\int d^d\xi \;\sqrt{\det\left({\tilde g}+R\sum_{i}^n F^{i}_{ab}\right)}
\eea
where 
$R$ is a typical scale of the theory ({\it e.g.} the radius of curvature).  An advantage of this definition is that, it is directly sensitive  to
the background charge. This is in contrast to the holographic entanglement entropy or Wilson loop
where the effects of the background charge is due to the metric components. 

For sufficietnly small charges one may expand the square  root which for $n=1$  and at leading order
one arrives at
\bea\label{rrr}
{\Gamma}=\frac{1}{G_{d+2}}\int d^d\xi \;\sqrt{\det({\tilde g})}\left(1-\frac{1}{4}R^2 F_{ab}^2\right),
\eea
which, in turns, shows that  in this limit it essentially contains the same information as the geometric
entropy, as we will explicitly demonstrate in the next section. 

For arbitrary charges, following the general idea of AdS/CFT correspondence, it is then natural to minimize $\Gamma$. The resultant quantity might be  used to define an order parameter
 which could probe different phases of the system as  we will demonstrate in the following sections, within  a specific model.

The paper is organized as follows. In the next section we will consider charge black branes with one $U(1)$ charge and
then compute the quantity \eqref{area}
 where we explore its different properties.
In section three we redo the same computations for the charge black hole in a  global AdS geometry. 
 The last section is devoted to  discussions.


\section{Electrically charged black brane solutions}

In  this section in order to explore  a possible information encoded in the expression defined by \eqref{area} we will consider a particular model consisting of the Einstein gravity with a negative 
cosmological constant coupled to a $U(1)$ gauge field. In this case the action \eqref{action}
reduces to
\be\label{RNaction}
I=\frac{1}{16\pi G_{d+2}}\int d^{d+2}x\;\sqrt{-g}\left({\bf R}-2\Lambda
-\frac{1}{4}F^2\right).
\ee
This model admits several  vacuum solutions which could be either
electric or dyonic   black branes (holes) charged under the $U(1)$ gauge field.  In what follows we 
will consider the electric case and will postpone the dyonic one to the section four.

Let us consider a $d+2$ dimensional (Euclidean) Reissner-Nordstrom  AdS black brane  solution which 
for $d\geq 2$   may be written as follows\cite{Chamblin:1999tk}
\footnote{Actually for $d=1$
we still have the same solution but with $f=1-r^2+\frac{Q^2}{2} r^2\ln r$ and $F_{rt}=\frac{Q}{r}$.}
\bea\label{RN}
&&ds^2=\frac{R^2}{r^2}\left(-f(r) dt^2+\frac{dr^2}{f(r)}+\sum_{i=1}^{d}dx_i^2\right),\;\;\;\;
F_{rt}=-QR  \sqrt{2d(d-1)}r^{d-2},\cr
&&f(r)=1-(1+Q^2 r_H^{2d})\left(\frac{r}{r_H}\right)^{d+1}
+Q^2 r^{2d},
\eea
where $R=\sqrt{-\frac{d(d+1)}{2\Lambda}}$ and $r_H$ are the radii of curvature and  horizon, respectively. The Hawking temperature in 
terms of the  radius of the horizon is 
\be
T=\frac{d+1 }{4\pi r_H}\left(1-\frac{d-1}{d+1} Q^2 r_H^{2d}\right).
\ee

This geometry is supposed to provide a gravitational description for a $d+1$ dimensional
CFT at finite temperature and density. The corresponding  chemical potential is 
\be
\mu=\sqrt{\frac{2d}{d-1}}Q Rr_H^{d-1}.
\ee

Let us consider the following strip as a subsystem in the dual  $d+1$ dimensional theory 
\be\label{strip}
0\leq t\leq \tau,\;\;\;\;-\frac{\ell}{2}\leq x_{d-1}\leq \frac{\ell}{2},\;\;\;\;0\leq x_i\leq L, \;\;\;\;\;x_d={\rm fixed}
\ee
for $i=1,\cdots,d-2$. Then there is a hypersurface in the bulk whose intersection with the boundary coincides with  the above strip. The profile  of the corresponding  hypersurface  
may be given by $x_{d-1}=x(r)$. Thus the induced (Euclidean) metric on the hypersurface is 
\be
ds^2_{\rm ind}={\tilde g}_{\mu\nu}dx^\mu dx^\nu=\frac{R^2}{r^2}\left[fdt^2+\left(\frac{1}{ f}+x'^2\right)dr^2+\sum_{i=1}^{d-2}dx_i^2\right],
\ee
where prime represents derivative with respect to $r$. In this case  the expression \eqref{area},   taking into account the solution 
 \eqref{RN} and the boundary subsystem \eqref{strip},  reads 
\be
{\Gamma}=\frac{\tau L^{d-2}R^d}{G_{d+2}}\int dr\; \frac{\sqrt{1-\phi^2+fx'^2}}{r^d}, 
\ee
where $\phi=\sqrt{2d(d-1)} Q r^{d}$.

Now the aim is to minimize $\Gamma$. Actually there is a standard procedure to minimize
$\Gamma$ by which the expression of $\Gamma$ may be 
treated as a one dimensional action for $x$ whose momentum conjugate is a constant of motion.
Therefore one arrives at 
\be\label{Con}
\frac{f x'}{r^d \sqrt{1-\phi^2+fx'^2}}=c,
\ee
where $c$ is a constant which can be fixed at a particular point. Usually the particular point is chosen to be the turning point where $x'\rightarrow \infty$ in which $x'$ drops from the
left hand side leading to a constant which is given in terms of a function of $r$ evaluated at the turning point.  When we are not explicitly considering the
effects of gauge field, {\it e.g.} in the computation of holographic entanglement or geometric  entropies where there is no $F$ in the square root, then the position of turning point 
is located between boundary and horizon. Whereas in the present case the situation is different. 

Actually as we shall see when we  increase the background charges the effects of gauge field
become important leading to a new scale in the theory which could take over the role of the horizon. More precisely, as it is evident from the equation \eqref{Con}, for a  given background charge 
there is a special  point at which $\phi=1$ that is given by
\be
r_\phi= \left(\frac{1}{2d(d-1) Q^2}\right)^{\frac{1}{2d}}.
\ee
Note that although at this point the  $x'$ dependence is dropped from the  left hand side of the equation \eqref{Con}, it  is not  a turning point. 
Moreover one can convince ourselves that 
the minimization  makes sense only  for $r\leq r_{\phi}$. In other words, in the present case the location of the turning point well be  between
boundary and $r_{\rm min}$ where $r_{\rm min}={\rm Min}(r_H,r_\phi)$, {\it i.e.} $0 \leq r_t \leq r_{\rm min}$, with
$r_t$  being the turning point. In what follows we will consider both 
$r_{\rm min}=r_H$ and $r_{\rm min}=r_{\phi}$ cases.

\subsection{$r_{\rm min}=r_H$ case}

Let us assume  $r_{\rm min}=r_H$ which happens if 
\be 
Q\leq Q_c=\frac{1}{\sqrt{2d(d-1)} r_H^{d}},\;\;\;\;{\rm or}\;\;\;\mu\leq \mu_c=\frac{R}{(d-1)r_H}.
\ee
In this case  one finds
\bea\label{ELL}
\ell=2\int_{0}^{r_t}dr \left(\frac{f_t}{f^2}\right)^{1/2}\left(\frac{r}{r_t}\right)^d
\frac{\sqrt{1-\phi^2}}{\sqrt{1-\left(\frac{r}{r_t}\right)^{2d}\frac{f_t}{f}}}.
\eea
 where $f_t=f(r_t)$. On the other hand using the equation \eqref{Con} one arrives at
\bea\label{GAMMA}
\Gamma=\frac{\tau L^{d-2}R^{d}}{G_{d+2}}\int_{\epsilon}^{r_t}dr 
\frac{\sqrt{1-\phi^2}}{r^d\sqrt{1-\left(\frac{r}{r_t}\right)^{2d}\frac{f_t}{f}}},
\eea
where $\epsilon$ is a UV cut off. From these expressions it is clear that there is a new  scale in the theory that  controls  the effects of the background filed, as we anticipated.
Of course since for the moment  we are in the range  of $Q\leq Q_c$, the new scale is irrelevant in what follows. We will back to the case 
of $Q> Q_c$ latter.

If one drops the factor of $\sqrt{1-\phi^2}$, the above expressions reduce to that of the geometric entropy studied in \cite{{Fujita:2008zv},{Bah:2008cj}}. Moreover
for pure  $AdS_{d+2}, d\geq 2 $ one has\cite{RT:2006PRL}
\bea\label{GEAdS}
\Gamma&=&\frac{\tau L^{d-2} R^d}{G_{d+2}}\bigg{[}\frac{1}{(d-1)\epsilon^{d-1}}-\frac{2^{d-1}\pi^{d/2}}{d-1}
\left(\frac{\Gamma\left(\frac{d+1}{2d}\right)}{\Gamma\left(\frac{1}{2d}\right)}\right)^d
\frac{1}{\ell^{d-1}}\bigg{]},
\eea
which is the expression of holographic entanglement entropy. Note also that for $d=1$ one gets a logarithmic behavior, $\Gamma\sim \ln\frac{\ell}{\epsilon}$. 

For the  RN background given in the equation \eqref{RN} we cannot find an analytic expression
for $\Gamma$ as a function of $\ell$. Nevertheless we can utilize a  numerical method to find
$\Gamma(\ell)$ numerically. This is, indeed, what we shall do in this subsection.  To proceed let us first explore  the behavior of $\ell$ as a function of $r_t$. 

From the expression \eqref{ELL} one finds that  for sufficiently small 
$r_t$ where $\Gamma$ probes the UV region of the theory the width $\ell$ vanishes
as $\ell\sim r_t\rightarrow 0$.  Moreover, in the opposite limit, 
the width $\ell$ also goes to zero as the turning  point approaches
the horizon. It is, indeed, due to the facts  that  $f_t\rightarrow 0$ for $r_t\rightarrow r_H$ and also 
 the integrand does not diverge faster than $1/f_t$. Therefore 
for $0\leq r_t\leq r_H$ the width $\ell$ goes to zero at both bounds and reaches a maximum
value in this interval.

This behavior can be
demonstrated  by solving the integral \eqref{ELL} numerically. To do so, by making use of a scaling,
without loss of generality, one may set $r_H=1$. Then the only parameter of the model is the 
charge of the solution.
 Note that  in this case one has  $0\leq Q^2\leq \frac{ 1}{2d(d-1)}$.  The neutral black brane corresponds to $Q=0$, while $Q^2=\frac{1}{2d(d-1)}$ is the case where $r_H=r_{\phi}$.  The behavior of $\ell$ as a function of $r_t$
for different values of $Q$ for $d=2$ are shown in the figure 1.
\begin{figure}
\begin{center}
\includegraphics[scale=.54]{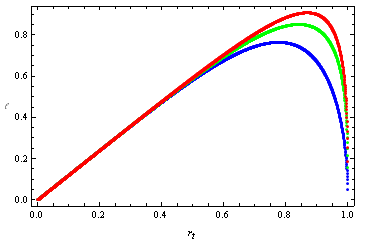}
\caption{ $\ell$  as a function of $r_t$ 
for the cases of $Q=0,0.3, 0.5$ which are shown by red, green and, blue, respectively. The blue curve corresponds to the case of $r_H=r_\phi$ while
the red one represents the neutral black brane. }
\end{center}
\end{figure}

Form the equation \eqref{ELL} one may, in principle, find the turning point as a function of $\ell$. Then
plugging the result into the equation \eqref{GAMMA} we get an expression for   $\Gamma$ as a function $\ell$.
It is important to note that since $\ell$ is not a one-to-one function of $r_t$ one has to make sure
that the resultant $\Gamma$ is minimum. Of course it is clear that the minimum $\Gamma$ is obtained
from the minimum $r_t$. 

It should also be noticed that  since the space time has a horizon one could always imagine the case where the function $\Gamma$ is
minimized by another hypersurface consisting of two disconnected parallel surfaces 
suspending between boundary and horizon. Therefore it is crucial to see which one is
smaller.

The disconnected solution is given by setting $r_t=r_H$ in the expression of $\Gamma$ by which we arrive at
\be
\Gamma^{\rm diss}=\frac{\tau L^{d-2}R^{d}}{G_{d+2}}\int_{\epsilon}^{r_{H}}dr 
\frac{\sqrt{1-\phi^2}}{r^d}.
\ee
 In general, depending on the parameters of the model  either of connected or disconnected solutions 
could be smaller. In order to compare these two solutions it is useful to define the difference between them
 as follows 
\bea\label{OP}
\Delta\Gamma&=&\Gamma^{\rm con}-\Gamma^{\rm dis}\\ &&\cr
&=&\frac{\tau L^{d-2}R^{d}}{G_{d+2}}\left[\int_{0}^{r_t}dr \left(
\frac{\sqrt{1-\phi^2}}{r^d\sqrt{1-\left(\frac{r}{r_t}\right)^{2d}\frac{f_t}{f}}}-
\frac{\sqrt{1-\phi^2}}{r^d}\right)
-\int_{r_t}^{r_{H}}dr \frac{\sqrt{1-\phi^2}}{r^d}
\right].\nonumber
\eea
Note that, although, both connected and disconnected  solutions are UV divergent, the UV contribution
drops out in the difference leading to a finite number. The behaviors of $\Delta\Gamma$ as a function
of $\ell$ for different  values of $Q$ for $d=2$ case have been drown in the figure 2.
\begin{figure}
\begin{center}
\includegraphics[scale=.54]{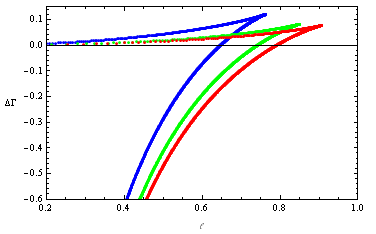}
\caption{$\Delta\Gamma$ as a function of $\ell$
for the cases of $Q=0,0.3, 0.5$ which are shown by red, green and blue, respectively. The blue curve corresponds to the case where $r_H=r_\phi$ 
and red one corresponds to the neutral case. Here we have set $r_H=1$  and $\frac{\tau L^{d-2}R^d}{G_{d+2}}=1$.}
\end{center}
\end{figure}

One observes that for sufficiently small $\ell$ the closed hypersurface minimizes the expression of
$\Gamma$, though there is a critical width over which the disconnected solution is favored.
Moreover the critical width is always smaller than the maximum value the width can reach. 
Therefore one may 
conclude that $\Gamma$ undegoes a sort of  a phase transition before it reaches the maximum $\ell$.  
 It is worth to note that 
as we increase the charge 
the maximum width becomes smaller and the phase transition occurs at smaller width, nevertheless as long as $Q\leq Q_c$ the behavior is universal which is that of geometric entropy.

Therefore as far as the qualitative behavior of $\Gamma$ is concerned the effects of gauge field are not important and the main contributions come
from the metric. In fact the background gauge field  only affects  the  position of the horizon.  We note, however, that as we increase the background charge
 one expects the effects of background charges become  important as we shall explore in the following subsection.

\subsection{$r_{\rm min}=r_{\phi}$ case}

To study  the effects of the background gauge field one may increase the background charge\footnote{
Since in our notations we have set $r_H=1$ there is an upper bound on the background charge. More precisely the allowed values of 
background charge is $ 1/(2d(d-1)) \leq Q^2 \leq (d+1)/(d-1)$. Note that $Q^2=(d+1)/(d-1)$ corresponds to the extremal case where $T=0$.  } so that  $Q>Q_c$
 where  $r_{\rm min}=r_{\phi}$. This  indicates that  the maximum value the turning point can get is $r_{\phi}$. More precisely one has  $0\leq r_t \leq r_{\phi} < r_H$.
 In other words, since the turning point cannot reach the horizon we will  not 
have the disconnected solution. 

Indeed looking at the equation \eqref{ELL}  one finds that although  the width vanishes  in the limit of  $r_t\rightarrow 0$,  it  terminates at a non-zero 
value as one approaches $r_\phi$. By making use of the numerical method the width $\ell$  can be found as a function of turning point which has been 
depicted in the figure 3 (left).

Moreover since in the present case we do not have the disconnected solution,  it does not make sense to compute the difference $\Delta\Gamma$.
 Indeed the
function $\Gamma$ is the quantity we may want to  compute. We note that due to the UV contribution, $\Gamma$ diverges and has to be regulated by a UV cut off. More 
precisely one gets
\be
\Gamma=\frac{\tau L^{d-2}R^{d}}{G_{d+2}}\int_{\epsilon}^{r_t}dr 
\frac{\sqrt{1-\phi^2}}{r^d\sqrt{1-\left(\frac{r}{r_t}\right)^{2d}\frac{f_t}{f}}}=\frac{1}{G_{d+2}}\frac{\tau L^{d-2}R^{d}}{(d-1)\epsilon^{d-1}}+\Gamma_{{\rm finite}}.
\ee
Subtracting the divergence part, it is then straightforward to calculate the finite part, $\Gamma_{\rm finite}$,  numerically. The results are shown in figure 3 (right).
\begin{figure}
\begin{center}
\includegraphics[scale=.41]{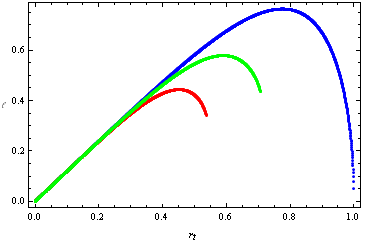}
\includegraphics[scale=.44]{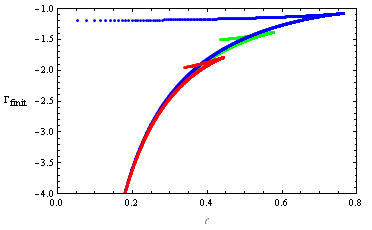}
\caption{$\ell$ and  $\Gamma$ as  functions of $r_t$and $\ell$ for  
$Q=0.5, 1,\sqrt{ 3}$ which are shown by blue, green and red, respectively. Note that $Q=0.5$ corresponds to the case of $r_H=r_\phi$ and we have plotted it just 
for a comparison. Here we have set $r_H=1$  and $\frac{\tau L^{d-2}R^d}{G_{d+2}}=1$.}
\end{center}
\end{figure}

From our numerical results one observes that as long as we are in the range of $Q_c<Q\leq \sqrt{(d+1)/(d-1)}$,  qualitatively the behavior of $\Gamma$ is universal, though
it decreases as one increases the charge. 
Indeed, there is a critical width, $\ell_c$ above that both $\Gamma$ and $\ell$ are not single valued functions. In other words for each width $\ell> \ell_c$ there 
are two turning points. Of course the favored $\Gamma$ corresponds to the smaller turning point. Moreover there is a maximum width over which there is no a closed hypersurface.
It is important to note that the width gets its maximum value before the turning point reaches its maximum value at $r_\phi$. 

An interesting observation we have made is as follows. Although there is a maximum width (or correspondingly a maximum turning point) over which there is no a 
closed hypersurface which minimizes $\Gamma$, there is a single closed hypersurface when $r_t=r_\phi$. Actually, as we have already mentioned, in this case  $r_t$ is not 
a turning point and indeed  the hypersurface can cross
the $r=r_\phi$ point and reaches the horizon. In fact  it is easy to see that for this case the horizon is a turning point. Therefore we will get 
a single distinctive closed hypersurface which can probe the charged  horizon while the effects of charges are important.  In this case the corresponding expressions
for $\ell$ and $\Gamma$ are given by
\bea
\frac{\ell}{2}=\int_{0}^{r_H}dr 
\frac{\left(\frac{f_\phi}{f^2}\right)^{1/2}\left(\frac{r}{r_\phi}\right)^d\sqrt{1-\phi^2}}{\sqrt{1-\left(\frac{r}{r_\phi}\right)^{2d}\frac{f_\phi}{f}}},\;\;
\Gamma=\frac{\tau L^{d-2}R^d}{G_{d+2}}\int_{\epsilon}^{r_H}dr 
\frac{r^{-d}\sqrt{1-\phi^2}}{\sqrt{1-\left(\frac{r}{r_\phi}\right)^{2d}\frac{f_\phi}{f}}},
\eea
where $f_\phi=f(r_\phi)$. In the figure 4 we have depicted 
the behaviors of $\ell$ and finite part of $\Gamma$ as functions of $Q$. 
\begin{figure}
\begin{center}
\includegraphics[scale=.46]{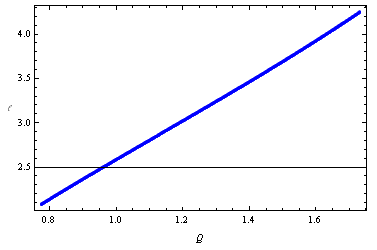}
\includegraphics[scale=.51]{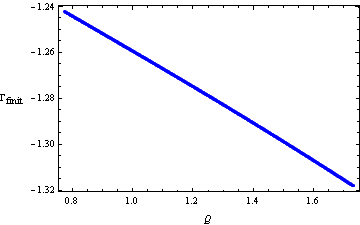}
\caption{$\ell$ and  $\Gamma_{{\rm finite}}$ as  functions of $Q$ for the case of $r_t=r_\phi$. The numerical values are 
for $r_H=1$ and  $R=L=1$. Note that for all values of $Q$ in the above plots we have $r_\phi < r_H$.}
\end{center}
\end{figure}
Note that as we increase the background charges the width also increases linearly though the finite part of $\Gamma$ decreases linearly.



\section{Black hole in global AdS}

In this section we extend  our study to a charged black hole in a global AdS geometry. The action 
is still given by \eqref{RNaction}. The corresponding $d+2$ dimensional charged black hole  may be written as\cite{Chamblin:1999tk}
\bea
&&ds^2=\frac{R^2}{r^2}\left(-f(r)dt^2+\frac{dr^2}{f(r)}+R^2d\Omega_{d}^2\right),\;\;\;\;\;\;
F_{rt}=-QR \sqrt{2d(d-1)}r^{d-2},\cr &&\cr
&&f(r)=1+\frac{r^2}{R^2}-\left(1+\frac{r_+^2}{R^2}+Q^2r_+^{2d}\right)\left(\frac{r}{r_+}\right)^{d+1}+
Q^2r^{2d},
\eea
where in our notation $d\Omega_{d}^2=d\theta^2+\cos^2\theta\;d\Omega_{d-1}^2$ with 
$d\Omega_{d-1}^2$ being the metric of a $(d-1)$-sphere, and $r_+$ is the location of the horizon which is a solution of $f(r)=0$. We note that 
in general $f=0$  has two real positive solutions and the  horizon is given by the smallest root. The Hawking temperature and 
chemical potential are 
\be
T=\frac{d+1}{4\pi r_+}\left[1-\frac{d-1}{d+1}\left(Q^2r_+^{2d}-\frac{r_+^2}{R^2}\right)\right],\;\;\;\;\;\mu=\sqrt{\frac{2d}{d-1}}Q R r_+^{d-1}.
\ee
Using the corresponding Euclidean action the phase space of this system has been studied in\cite{Chamblin:1999tk} where  it was shown that the theory has a rich phase space.
Indeed the system could  be thought of as either a grand canonical ensemble or a canonical ensemble  depending on whether one wants to keep
chemical potential or electric charge fixed, respectively. In either cases there are a critical values for the parameters over which the model exhibits different behaviors.

Holographic geometric entropy in this  background has also  been studied \cite{Bah:2008cj} where it was shown that it may provide a useful
order parameter to probe different phases of the system. Note that since in this case one, usually, performs  a double Wick rotation there are two
different ways to embed the hypersurface in the bulk. One could either consider $r(t)$ or $r(\theta)$.
Actually by making use of these embedding it was observed  in \cite{Bah:2008cj}  that the resultant phase structures are very 
similar to that  obtained from the  Euclidean action\cite{Chamblin:1999tk}. We note, however, that  since in what follows we are 
interested in the effects of the gauge field, as defined in the equation \eqref{area},  the $r(t)$ embedding  should automatically be excluded.

Therefore  we will consider a subsystem in the form of  $S^{d-2}\times \mathbb{R}\times I$, with $I$ being  an interval along $\theta$ direction given by 
$0\leq \theta\leq 2\pi \frac{\ell}{R}$ with $\ell <R$. The extension of this
subsystem to the bulk leads to a hypersurface  whose profile is given by $\theta=\theta(r)$. Thus
the  induced (Euclidean) metric on the hypersurface is
\be
ds^2=\frac{R^2}{r^2}\left[f dt^2+\left(\frac{1}{f}+R^2\theta'^2\right)dr^2+R^2\cos^2\theta\;d\Omega_{d-2}^2\right]
\ee
Therefore we arrive at 
\be\label{rr}
\Gamma=\frac{\tau V_{d-2} R^{d}}{G_{d+2}}\int_\epsilon^{r_t} dr \;\frac{\cos^{d-2}\theta}{r^d}
\sqrt{1-\phi^2+f R^2\theta'^2},
\ee
where $V_{d-2}$ is the volume of $(d-2)$-sphere with radius $R$ and $r_t$ is the turning point
where $\theta'(r)$  diverges.

Alternatively, for $d\geq 3$ one may use a notation in which 
\be
d\Omega_d^2=d\psi^2+\cos^2\psi\; d\theta^2+\sin^2\psi (d\phi^2+\cos^2\phi d\Omega^2_{d-3}),
\ee
and thus the corresponding subsystem may be chosen so that $\phi={\rm constant}$. The constant may be set to  $\phi=\pi/2$ and the profile of the hypersurface is 
given by $\psi(r)$. Therefore the induced (Euclidean) metric is 
\be
ds^2=\frac{R^2}{r^2}\left[f dt^2+\left(\frac{1}{f}+R^2\cos^2\psi\; \theta'^2\right)dr^2+R^2 d\psi^2+R^2\sin^2\psi\;d\Omega_{d-3}^2\right].
\ee
So that
\be\label{vv}
\Gamma=\frac{\tau V_{d-3} R^{d}}{G_{d+2}}\int_\epsilon^{r_t} d\psi\; dr \;\frac{\cos^{d-3}\psi}{r^d}
\sqrt{1-\phi^2+f R^2\cos^2\psi\;\theta'^2}.
\ee

Now the aim is to minimize $\Gamma$  given  in the equation \eqref{rr} or \eqref{vv},  which can be done by treating them as actions for $\theta$. 
In what follows we will mainly consider the first case where $\Gamma$ is given by the equation \eqref{rr} where  unlike the previous cases, except for $d=2$, 
the momentum conjugate of $\theta$ is not
a constant of motion and therefore one needs to directly solve the equation of $\theta$  which is
\be\label{diffeq}
\frac{d}{dr}\left(\frac{\cos^{d-2}\theta}{r^d}\;\frac{f R^2\theta'}{\sqrt{1-\phi^2+f R^2\theta'^2}}\right)
+(d-2)\sin\theta\;\cos^{d-3}\theta\;\frac{\sqrt{1-\phi^2+f R^2\theta'^2}}{r^d}=0.
\ee
This equation may be solved with proper boundary conditions to find $\theta$ as a function of $r_t$. The corresponding boundary conditions could be
$\theta(r\rightarrow 0)=2\pi \frac{\ell}{R}$ and $\theta(r_t)=0$.
Then plugging the result into  the equation \eqref{rr} one can find $\Gamma$ as a function of $\ell$. 

Although it is not explicitly  clear from the above equation, there is still a special point at  $r=r_\phi$ where $\phi=1$ and 
 the minimization makes sense for $r\leq r_\phi$. Indeed the situation is very similar to what we have considered in
the previous section for the black brane. In particular for $r_+\leq r_\phi$ 
the function  $\Gamma$ may also be minimized by  a disconnected hypersurface which in the  present case is given by 
\be
\Gamma^{\rm diss}=\frac{ V_{d-2} R^{d-1}}{G_{d+1}}\int_\epsilon^{r_{H}} dr \;\frac{\cos^{d-2}\theta_0}{r^d}
\sqrt{1-\phi^2},
\ee
where $\theta_0=\theta(r=0)$. It is then natural to look for $\Delta\Gamma$ as a function of 
$\ell$.

To proceed let us first consider   $d=2$ case in which the momentum conjugate of $\theta$ is, indeed,
a constant of motion
\be
\frac{ R\theta'}{\sqrt{1-\phi^2+f R^2\theta'^2}}=\left(\frac{r}{r_t}\right)^d\;\frac{f_t^{1/2}}{f},
\ee
where $r_t$ is the turning point. so that 
\bea
\ell=\frac{1}{\pi}\int_{0}^{r_t}dr
\frac{ \left(\frac{f_t}{f^2}\right)^{1/2}\left(\frac{r}{r_t}\right)^2\sqrt{1-\phi^2}}{\sqrt{1-\left(\frac{r}{r_t}\right)^{4}\frac{f_t}{f}}},\;\;\;
\Gamma=\frac{\tau R^2}{G_{4}}\int_{\epsilon}^{r_t}dr 
\frac{\sqrt{1-\phi^2}}{r^2\sqrt{1-\left(\frac{r}{r_t}\right)^{4}\frac{f_t}{f}}},
\eea
which have essentially the same form as the corresponding expressions we have found  in the previous section, though the function $f$ is
different.  Therefore one expects that the system may exhibit the same behavior as in the
black brane. In particular one can show that as long as we are in the range of the parameters where $r_+\leq r_\phi$, the corresponding width, $\ell$, vanishes at
 both  $r_t=0$ and $ r_t=r_+$ points, while for $r_\phi < r_+$ although the width  vanishes at $r_t$, it tends  to a non-zero constant as $r_t\rightarrow r_\phi$. 
Moreover $r_t=r_\phi$  is not a turning point and the hypersurface can cross the point of $r=r_\phi$ to reach the 
horizon which is, indeed, the turning point in this case. 

 In order to  calculate $\Gamma$ one  distinguishes  two different 
cases depending on whether $r_+\leq r_\phi$ or $r_+> r_\phi$. Indeed for sufficiently small charges, {\it i.e.} $Q\leq Q_c$,  where we are in the region of $r_+\leq r_\phi$  the main 
contributions come from the  metric 
and the effects of the charge is only due to the location of the horizon which is encoded in the metric's components.
 Indeed in this case the behavior of $\Gamma$ is the same as the  holographic geometric entropy.

 On the other hand as one increases the background charge so that    $Q> Q_c$ one reaches the region  $r_\phi < r_+$ where  
 the effects of the background charge become  important. In this region since  there
is no place where  the hypersurface  can end,  the minimization procedure does not lead to the disconnected solution. 

It is worth to mention that for $d\geq 3$ using the expression of $\Gamma$ given in the equation \eqref{vv} we get exactly the same behavior as that in $d=2$
discussed above  which is, indeed,  the same as what we have found  
in the previous section displayed in the figures 1 and 3.

On the other hand using the expression \eqref{rr} for $d\geq 3$ although qualitatively we get the same behavior, a new feature appears when we 
change the ratio of $R/r_+$. Of course as  far as the effect of the gauge field is concerned the situation remains unchanged. Namely $\phi=1$ sets
a scale which controls the effects of the gauge field as before. 

In order to explore the new feature let us consider the situation where  $0\leq Q^2\leq \frac{1}{2d(d-1)}$ which corresponds to the case of  $r_+\leq r_\phi$.
Note  that in this region the effect of the background field is irrelevant and indeed we could have done the same for the geometric entropy. To proceed  it is useful to study the
behavior of  $\Delta \Gamma$ which we will do that by using  a numerical method. By making use of a  scaling one may set $r_+=1$. It is important to note that unlike  
the black brane case where $\Delta \Gamma$ depends on  $R$ just through a trivial overall factor, in the present case it appears in the function $f$ and therefore 
it may affect the behavior of the order parameter. To find the corresponding behavior numerically we will fix the dimension and the charge, and therefore
we are left with a  free parameter $R$ which controls the behavior of the order parameter. Indeed one observes that for $R$ of order of $r_+$ or bigger the model
undergoes a phase transition though for a sufficiently  small $R/r_+$ it exhibits no phase transition. More precisely there is a critical $R/r_+$ that 
indicates whether the system exhibits a phase transition.
 In the figure 5 we have summarized the above discussions by  plotting $\Delta \Gamma$ as a function of $\ell$ for different values of 
$R$.\begin{figure}
\begin{center}
\includegraphics[scale=.50]{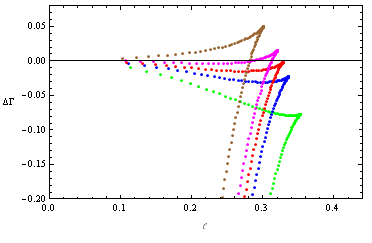}
\caption{
$\Delta\Gamma$ as a  function of $\ell$  for  $Q=0.3, r_+=1$ and,  $R= 0.6, 0.66, 0.69, 0.72, 0.8$ which are shown by
green, blue, red, magenta and brown respectively. Note that as far as $0\leq Q\leq 0.5$ where $r_+\leq r_\phi$ one gets qualitatively the
same behavior.}
\end{center}
\end{figure}



\section{Discussions}

In this paper we have introduced a quantity which is sensitive to the background fractionalized charge not only due to its effects in the components of the metric, but
also directly from the  gauge field. To explore its  properties we have explicitly computed the quantity for 
the RN black-brane and a black-hole  in an asymptotically 
AdS geometry.

For sufficiently small charges the metric plays the essential roles, while as one increases the charge one would expect to see the effects of the 
gauge field. Indeed following our definition  in the quantitie \eqref{area},  there is natural scale over which the direct effects of gauge 
field become significant. To elaborate this point it is illustrative  to study the induced metric in more detail. To proceed it is  useful to recall  the following identity 
\be
\sqrt{\det\left({\tilde g}+R{F_{ab}}\right)}=\bigg[\det(\tilde{g})\det(G)\bigg]^{1/4},
\ee
where
\be
G_{\mu\nu}={\tilde g}+R^2F_{\mu\rho}{\tilde g}^{\rho\sigma}F_{\sigma\nu}.
\ee
 In our case, using the explicit expression for $x'$ obtained, for example, from the equation \eqref{Con}  the
induced (Euclidean) metric may be recast to the following form
\be
ds_{\rm ind}^2=\frac{R^2}{r^2}\left[f dt^2+\left(\frac{f-f_t\phi^2\;(\frac{r}{r_t})^{2d}}
{f-f_t\; (\frac{r}{r_t})^{2d}}\right)\;
\frac{dr^2}{f}+\sum_{i=1}^{d-2} dx_i^2\right],
\ee
which shows that there is a horizon at $r=r_H$, as expected. On the other hand for the metric $G_{\mu\nu}$ one
finds
\be\label{mopen}
ds_{\rm open}^2=\frac{R^2}{r^2}\left\{
\frac{f\;(1-\phi^2)}{f-f_t\phi^2\;(\frac{r}{r_t})^{2d}}
\left[f dt^2+\left(\frac{f-f_t\phi^2\;(\frac{r}{r_t})^{2d}}{f-f_t\; (\frac{r}{r_t})^{2d}}\right)\;
\frac{dr^2}{f}\right]+\sum_{i=1}^{d-2} dx_i^2\right\},
\ee
that indicates a possibility of having a natural scale  at  $r=r_\phi$ where $\phi=1$, though the original geometry is smooth at this point.
Indeed for $Q\leq Q_c$ one has  $r_H\leq r_\phi$. Therefore  the scale $r_\phi$ is behind the horizon  and does not play an essential role
indicating that the main contributions come from the metric. In fact in this case the effect of the charge is only through the components of the metric which in turn may 
fix the position of the horizon and indeed  qualitatively the function $\Gamma$ has the same
behavior as the geometric entropy.

On the other hand in the opposite  limit when  $Q>Q_c$ where one has  $r_\phi\leq r_H$ the effect of background gauge field is important and 
for a generic value of $r_t$ the solution is well defined if  $0\leq r_t < r_\phi$.  Note that in this case one gets a`` bubble solution'' and therefore the horizon cannot 
be probed. In this case the behavior of the function $\Gamma$ still is qualitatively the same as 
the geometric entropy, though since we are in the large charge limit, for fixed $r_H$, the corresponding dual theory
should be at low tempearture and therefore it does not exhibits a phase transition.

Note also that for the special value of $r_t=r_\phi$  the metric \eqref{mopen} is well defined at  $r=r_\phi$ and, indeed, it have a horizon at $r=r_H$.

Probably the most interesting, but rather difficult, aspect of our  study, is to find an interpretation for the quantity defined by the  equation \eqref{area} from 
dual field theory point of view.   Of course we should admit that we do not have a good answer
to this queation and indeed in this paper we have considered this quantity as a parameter which 
could probe the system. It would be very interesting to find the corresponding interpretation 
from field theory point of view.


\section*{Acknowledgments}

We would like to thank A. Davodi,  M. M. Mohammadi Mozaffar,  A. Mollabashi, A. E. Mosaffa,  M. R. Tanhayi and  A. Vahedi for useful discussions.
We would also like to thank D. Tong for comments on the draft of the paper.




\begin{thebibliography}{10}

\bibitem{M:1997}

J. M. Maldacena,
"The large N limit of superconformal field theories and supergravity,''
Adv.\ Theor.\ Math.\ Phys.\  {\bf 2}, 231 (1998)
[Int.\ J.\ Theor.\ Phys.\  {\bf 38}, 1113 (1999)]
[hep-th/9711200].

\bibitem{Witten:1998zw} 
  E.~Witten,
  ``Anti-de Sitter space, thermal phase transition, and confinement in gauge theories,''
  Adv.\ Theor.\ Math.\ Phys.\  {\bf 2}, 505 (1998)
  [hep-th/9803131].



\bibitem{Hartnoll:2011}
S. A. Hartnoll, ``Horizons, holography  and condensed matter,'' arXiv: 1106.4324.


\bibitem{SSV:2002}
  T. Senthil, S. Sachdev and M. Vojta, ``Fractionalized Fermi liquids,''
     Phys. \ Rev.  \ Lett.  {\bf 90} , 216403 (2003)  [arXiv: cond-mat/0209144]

\bibitem{Sachdev:2010um} 
  S.~Sachdev,
  ``Holographic metals and the fractionalized Fermi liquid,''
  Phys.\ Rev.\ Lett.\  {\bf 105}, 151602 (2010)
  [arXiv:1006.3794 [hep-th]].

\bibitem{Huijse:2011hp} 
  L.~Huijse and S.~Sachdev,
  ``Fermi surfaces and gauge-gravity duality,''
  Phys.\ Rev.\ D {\bf 84}, 026001 (2011)
  [arXiv:1104.5022 [hep-th]].




\bibitem{Hartnoll:2012ux} 
  S.~A.~Hartnoll and D.~Radicevic,
  ``Holographic order parameter for charge fractionalization,''
  Phys.\ Rev.\ D {\bf 86}, 066001 (2012)
  [arXiv:1205.5291 [hep-th]].

\bibitem{RT:2006PRL}
S.~Ryu and T.~Takayanagi,
"Holographic Derivation of Entanglement Entropy from AdS/CFT,''
Phys. Rev. Lett. {\bf 96} (2006) 181602
[hep-th/0603001].

\bibitem{Rey:1998ik} 
  S.~-J.~Rey and J.~-T.~Yee,
  ``Macroscopic strings as heavy quarks in large N gauge theory and anti-de Sitter supergravity,''
  Eur.\ Phys.\ J.\ C {\bf 22}, 379 (2001)
  [hep-th/9803001].


\bibitem{Maldacena:1998im} 
  J.~M.~Maldacena,
  ``Wilson loops in large N field theories,''
  Phys.\ Rev.\ Lett.\  {\bf 80}, 4859 (1998)
  [hep-th/9803002].




\bibitem{Fujita:2008zv} 
  M.~Fujita, T.~Nishioka and T.~Takayanagi,
  ``Geometric Entropy and Hagedorn/Deconfinement Transition,''
  JHEP {\bf 0809}, 016 (2008)
  [arXiv:0806.3118 [hep-th]].

\bibitem{Bah:2008cj} 
  I.~Bah, L.~A.~Pando Zayas and C.~A.~Terrero-Escalante,
  ``Holographic Geometric Entropy at Finite Temperature from Black Holes in Global Anti de Sitter Spaces,''
  Int.\ J.\ Mod.\ Phys.\ A {\bf 27}, 1250048 (2012)
  [arXiv:0809.2912 [hep-th]].

\bibitem{Aharony:2003sx} 
  O.~Aharony, J.~Marsano, S.~Minwalla, K.~Papadodimas and M.~Van Raamsdonk,
  ``The Hagedorn - deconfinement phase transition in weakly coupled large N gauge theories,''
  Adv.\ Theor.\ Math.\ Phys.\  {\bf 8}, 603 (2004)
  [hep-th/0310285].



\bibitem{Chamblin:1999tk} 
  A.~Chamblin, R.~Emparan, C.~V.~Johnson and R.~C.~Myers,
  ``Charged AdS black holes and catastrophic holography,''
  Phys.\ Rev.\ D {\bf 60}, 064018 (1999)
  [hep-th/9902170].



\bibitem{Basu:2005pj} 
  P.~Basu and S.~R.~Wadia,
  ``R-charged AdS(5) black holes and large N unitary matrix models,''
  Phys.\ Rev.\ D {\bf 73}, 045022 (2006)
  [hep-th/0506203].



\bibitem{Yamada:2006rx} 
  D.~Yamada and L.~G.~Yaffe,
  ``Phase diagram of N=4 super-Yang-Mills theory with R-symmetry chemical potentials,''
  JHEP {\bf 0609}, 027 (2006)
  [hep-th/0602074].





\bibitem{Harmark:2006di} 
  T.~Harmark and M.~Orselli,
  ``Quantum mechanical sectors in thermal N=4 super Yang-Mills on R x S**3,''
  Nucl.\ Phys.\ B {\bf 757}, 117 (2006)
  [hep-th/0605234].

 
\end{thebibliography}
\end{document}